\documentclass[useAMS,usenatbib]{mnras}

\usepackage{natbib, aas_macros}
\usepackage{ulem}
\usepackage[dvips]{graphicx}
\usepackage{wrapfig}
\graphicspath{{./figs/}}
\usepackage{color}
\usepackage{amsmath}
\usepackage{amssymb}
\usepackage{bm}

\newcommand{\Msun}{{\rm  M_{\odot}}}

\newcommand{\Zsun}{Z_{\odot}}

\newcommand{\HII}{$\rm{H_{II}}$ }
\newcommand{\Mh}{M_{\rm h}}

\newcommand{\Mbh}{{\rm M_{\rm BH}}}

\newcommand{\Msunyr}{{\rm \Msun~{\rm yr^{-1}}}}

\newcommand{\Rvir}{R_{\rm vir}}

\title[Gas and metal outflow from massive galaxies] 
{FOREVER22: Gas and metal outflow from massive galaxies in protocluster regions}
\author[Harada et al. ]{Naoki Harada$^{1}$\thanks{E-mail: nharada@bd.mbn.or.jp},Hidenobu Yajima$^{1}$ and Makito Abe$^{1}$ \\
$^{1}$Center for Computational Sciences, University of Tsukuba, Ten-nodai, 1-1-1 Tsukuba, Ibaraki 305-8577, Japan\\
}

\begin{document}

\date{Accepted 2023 August 22. Received Received 2023 August 21; in original form 2022 October 09}

\pagerange{\pageref{firstpage}--\pageref{lastpage}} \pubyear{2023}

\maketitle

\label{firstpage}

%
%
\begin{abstract}
We study gas and metal outflow from massive galaxies in protocluster regions at $z=3-9$ by using the results of the FOREVER22 simulation project. Our simulations contain massive haloes with $M_{\rm h} \gtrsim 10^{13}~\rm M_{\odot}$, showing high star formation rates of $> 100~\rm M_{\odot}~yr^{-1}$ and hosting supermassive black holes with $M_{\rm BH} \gtrsim 10^{8}~\rm M_{\odot}$.
We show that the mass loading factor ($\eta_{\rm M}$) sensitively depends on the halo mass and it is $\eta_{\rm M} = 1.2~(9.2)$ for $\Mh = 10^{13}~(10^{11})~\Msun$. 
Once the halo mass exceeds $\sim 10^{12.5}~\rm \Msun$, the outflow velocity of the gas rapidly decreases near a virial radius, and the gas returns to a galactic centre finally as a fountain flow.
Also, the metal inflow and outflow rates sensitively depend on the halo mass and redshift. At $z=3$, the inflow rate becomes larger than the outflow one if $\Mh \gtrsim 10^{13.0}~\Msun$. 
Thus, we suggest that massive haloes cannot be efficient metal enrichment sources beyond virial radii that will be probed in future observations, e.g., studies of metal absorption lines with the Prime Focus Spectrograph on the Subaru telescope.
\end{abstract}

%
%

%
%
\section{Introduction}
In the modern theory of cosmological structure formation, galaxies evolve via mergers and matter accretion from the intergalactic medium (IGM). 
The growth rate of galaxies sensitively depends on the environment in the large-scale structures. 
In the overdense regions, the matter accretion rate and merger frequency are expected to be larger than the cosmic mean fields. 
Thus, multiple massive/bright galaxies can collectively form in such overdense regions even in the early Universe. 
According to the hierarchical structure formation scenario, such particular regions finally evolve into local galaxy clusters \citep{Overzier16}. 
Therefore, the clustering regions of massive galaxies in the early Universe are dubbed protoclusters (PCs). 
Thanks to the brightness of massive galaxies in PCs, recent galaxy surveys have discovered a lot of PCs and allowed us to study the galaxy evolution in overdense regions 
\citep[e.g.,][]{Matsuda05, Toshikawa18, Higuchi19, Umehata19, Miller18, Hill20}.  
These observations indicate that a diversity of galaxy populations emerge in PC regions. However, the origin of the diversity has not been understood well. 

It is believed that the baryon cycle, i.e., accretion of matter, outflow, and recycling of gas, has an inseparable relationship with the evolution of galaxies. 
Star formation activities in galaxies are closely related to the supply of the gas accreted from circumgalactic medium (CGM) and IGM, and from galaxy mergers. 
The interstellar gas and metals can be returned into CGM/IGM via stellar/active galactic nuclei (AGN) feedback. 
Therefore, the physical states of CGM/IGM are tightly related to the activities of the galaxies. 
Some parts of the ejected components would fall back again into the galaxy due to its deep gravitational potential. 
This recycling process can affect the subsequent star formation histories. 
One of the central issues in galactic astronomy is to correctly follow the baryon cycle with the stellar/AGN feedback. 
So far, various theoretical works have proposed the numerical feedback models to reproduce the baryon cycle through the simulation of the galaxy formation \citep[e.g.,][]{Springel03, Schaye15, Yajima17c, Pillepich18}.
However, there is a large room in parameters for the star formation and the feedback models  in cosmological simulations of galaxy formation. 
By introducing the models and parameters appropriately, previous studies
have successfully reproduced observational properties, e.g.,  
 the star formation history 
\citep[e.g.,][]{Springel03, Oppenheimer06, Pillepich18}, the galaxy 
luminosity and mass function \citep[e.g.,][]{Benson03, Oppenheimer10, Schaye15, Pillepich18}, 
the galaxy size \citep[e.g.,][]{Pillepich18}, the mass-metallicity relation (MZR) \citep[e.g.,][]{Finlator08, 
Rossi17}, and the presence of metals in the diffuse IGM \citep[e.g.,][]{Aguirre01, Oppenheimer06}. 

Recent observations have been studying the baryon recycling process between galaxies, circumgalactic medium (CGM) and IGM, 
that can provide hints to understand the feedback mechanism related to the enhancement and quenching of star formation \citep[]{Tumlinson17}. 
\citet{Bertone07} introduced the model of recycling in their semi-analytical studies. 
Also, a lot of theoretical studies have been devoted to clear the relation between the baryon recycle and the galaxy evolution, for example, in the viewpoint of gas dynamics in CGM \citep{Oppenheimer08, Oppenheimer10, Ford14, Christensen16, Hafen19, Borrow20, Lochhaas20, Mitchell20, Hafen20, Wright21, Mitchell22}, the impacts of stellar and AGN feedback \citep{Shen13, Tollet19, Wright20, Donahue22}, stellar mass and star formation rates (SFRs) \citep{Dave11}, metal enrichment history \citep{Brook14, Muratov17}, and galaxy mass assembly \citep{Alcazar17, Mitchell22}. Among them, the particle tracking method in smoothed-particle hydrodynamics (SPH) simulations is a promising way to study and visualize the trace of gas flow in CGM and/or IGM \citep{Oppenheimer08, Oppenheimer10, Ford14, Christensen16, Hafen19, Borrow20}.
However, the mass range of galaxies investigated is still limited. Recent discoveries of high-redshift PC regions and passive galaxies bring the motivation of understanding the feedback and the outflow of very massive galaxies. 

Progenitors of today's high-mass "Coma-type" clusters with $\Mh > 10^{15}~\Msun$ is thought to have the halo 
mass of $\Mh > 10^{13}~\Msun$ at $z \sim 3$ \citep[]{Chiang13}. Simulations of massive PCs are  
challenging because of their rareness in space which requires large simulation volumes. In this paper, we perform cosmological SPH simulations with a large simulation box of $(714.3 ~\rm cMpc)^{3}$, and study the baryon cycling in massive dark matter (DM) haloes at redshifts of $3-9$ that are supposed to be progenitors of massive galaxy clusters in the local Universe. 

In Section 2, we explain FOREVER22 project and the star formation, and the feedback models. 
The inflow and outflow of gas in massive galaxies are presented in Section 3. 
We show the metal outflow rate and spatial distribution around massive haloes in Section 4. 
In Section 5, we summarize our main results.


%
%
\section{Simulation}
\subsection{The FOREVER22 simulation}
In this study, we use the result of FOREVER22 project which is a new series of cosmological hydrodynamic simulations focusing on 10 PC regions \citep{Yajima21}. 
We utilize the modified version of smoothed particle hydrodynamics (SPH)/$N$-body code GADGET-3 \citep{Springel01, Springel05e}. 
The code incorporates the star formation, the radiation feedback and supernova (SN) feedback from the stars, and the AGN feedback from the massive black holes (BHs). 
Here we use a standard SPH scheme that can suppress the mixing of clouds and background gas \cite[see also][]{Agertz07, Schaller15, Braspenning22}. Therefore, the outflow gas can propagate more easily. On the other hand, we carefully tune the subgrid models to be fit for use in the cosmological simulation with the standard SPH. 
Here we summarize the implementation of the code. 
More details of the simulations are described in \citet{Yajima21, Yajima22b}.

\subsection{Star formation}
In the FOREVER22 project, the star formation is followed by using the model developed in \citet{Schaye08} that is based on the Kennicutt-Schmidt law $\dot{\Sigma}_\ast = A(\Sigma_{\rm g}/1~\Msun~{\rm pc}^{-2})^n$ of local galaxies. 
The star formation rate $\dot{m}_\ast$ is evaluated from the total pressure $P$ of the interstellar medium (ISM) as follows 
\begin{equation}
    \dot{m}_{\ast} = m_{\rm SPH}A(1~\Msun~{\rm pc}^{-2})^{-n}\left(\frac{\gamma}{G}f_{\rm g}P\right)^{(n-1)/2},
\end{equation}
where $m_{\rm SPH}$ is the SPH particle mass, $\gamma = 5/3$ is the ratio of specific heats, $f_{\rm g}$ is the gas mass fraction (set to unity), $n$ and $A$ are the free parameter. 
During the simulation time step $\Delta t$, the conversion probability from an SPH particle to a stellar particle is evaluated by \citep{Schaye08}
\begin{equation}
    \dot{p}_{\ast} = \min \left(\frac{\dot{m}_{\ast} \Delta t}{m_{\rm SPH}},1\right). 
\end{equation}
In the simulation, the stellar particle is treated as a star cluster assuming a Chabrier initial mass function (IMF) with the mass range of $0.1 - 100~\Msun$.  
 We confirmed that the star formation and feedback models of the FOREVER22 project reproduced observational  properties, e.g., the cosmic star formation rate density and the stellar mass function \citep{Yajima21}, and most parameters  referred to the values derived in \citet{Schaye15}. 
The slope $n=1.4$ and $A=1.5\times10^{-4}~\Msunyr~{\rm kpc}^{-2}$ are the fiducial values, but the slope $n$ is set to 2.0 if the gas density exceeds the metallicity dependent critical value $n_0(Z/0.002)^{-0.64}~{\rm cm}^{-3}$. 
Also, if the gas density satisfies $n_{\rm H} > n_0$, the  effective equation of state with the adiabatic index $\gamma_{\rm eff}$ is used. 
To prevent the undesirable gravitational fragmentation in the unresolved scale, we set $\gamma_{\rm eff} = 4/3$. 
The critical density $n_0$ and the floor temperature at $n_{\rm H} = n_0$ are the parameters, which depend on the resolution of the simulation \citep[see \S 2.1 in][]{Yajima21}.
Here we set $n_{\rm 0}=0.1~\rm cm^{-3}$. 

The gas temperature is estimated with a net cooling rate derived based on the collisionally and photoionization equilibrium states of each gas and metal species using {\sc cloudy} v07.02 \citep{Ferland00}.
We take into account the line cooling processes of metals and primordial gas and the cooling by continuum radiation such as from thermal bremsstrahlung and recombination.

\subsection{Stellar feedback}
Young stars have an impact on the physical states of the interstellar medium (ISM), the CGM, and the IGM via radiation and SN feedback. 
These feedback processes are considered in our simulations as follows. 

\subsubsection{Radiation feedback}

First, the simulations take into account the photoionization feedback. 
We consider UV radiation only from star clusters with an age younger than $10^{7}~\rm yr$.
The photoionized region is evaluated by taking the balance between the ionizing photon production rate $\dot{N}_{\rm ion}$ of a stellar particle and the recombination rate of neighbour gas particles as
\begin{equation}
    \dot{N}_{\rm ion} = \sum_i n_{{\rm H},i}^{2}\alpha_{\rm B}\frac{m_i}{\rho_i}, 
\end{equation}
where $n_{{\rm H},i}$ $m_i$ and $\rho_i$ are hydrogen number density, mass and  density of $i$-th SPH particle, $\alpha_{\rm B}$ is the case-B recombination coefficient \citep{Hui97}. 
The local recombination rate is summed up in order of the distance from the stellar particle until the recombination rate reaches  $\dot{N}_{\rm ion}$. 
The $\dot{N}_{\rm ion}$ of each stellar particle is evaluated by using a synthesized SED of the Chabrier IMF considering its age and metallicity. 
Once the gas is ionized, the temperature of the photoionized SPH particles is set to be $3\times10^4~{\rm K}$ as hot \HII bubbles, and the star formation is forbidden in the \HII region. 
This feedback can affect the state of ISM if we have sufficient resolution to resolve the cold gas component. 
Therefore, this feedback can play an important role in the suppression of star formation significantly in the high-resolution series of the FOREVER22 simulations, First0 and First1 runs. 

In addition to the photoionization process, the UV photon absorption by dust gives momentum to gas. 
The simulation models the feedback of the UV radiation pressure by giving the momentum kick to the SPH particle. 
We evaluate the mean free path of the UV continuum photons against the dust as $l_{\rm mfp} = 1/\kappa_{\rm d}\rho$ where $\kappa_{\rm d}$ is the opacity of dust. 
If gas particles are within $l_{\rm mfp}$ from a star cluster, they are influenced by the radiation pressure as
\begin{equation}
    \bm F_{\rm rad} = \frac{\kappa_{\rm d} \rho L_{\rm UV}}{4\pi r^2 c}\frac{\bm r}{r}, 
\end{equation}
where $r$ is the distance from the stellar particle to the gas particle, $L_{\rm UV}$ is the luminosity for the UV continuum ($1000-5000~$\AA).

Besides, we take into account the UV background radiation at $z \lesssim 10$ by using the model of \citet{Haardt01}. 
Once the UV background turns on, the chemical compositions and the cooling rates can change due to the photo-ionization process. We consider the modifications of the cooling rates for 11 elements (H, He, C, N, O, Ne, Mg, Si, S, Ca, and Fe) under the UV background following \citet{Wiersma09a}. 
Also, the self-shielding effect for dens gas is incorporated. We assume that if the hydrogen number density is larger than $0.01~\rm cm^{-3}$, the flux decreases with a fraction $(n_{\rm H}/0.01~{\rm cm^{-3}})^{-2}$
\citep{Yajima12h, Rahmati13a}. Thus, the UV background suppresses the cooling of the low-density gas.

\subsubsection{Supernova Feedback}
The SN feedback is modelled by using the  recipe proposed by \citet{DallaVecchia12}. 
In this scheme, the SN energy is injected into neighbour gas particles as thermal energy. 
To avoid the over-cooling problem, the heating gas particles are stochastically chosen and are heated to $10^{7.5}~{\rm K}$. 
\citet{DallaVecchia12} argued that the injected thermal energy is converted into kinetic one if the cooling time is longer than the local sound crossing time. 
Thus, the SN feedback effectively works if the gas density is less than the following critical value, 
\begin{equation}
    n_{\rm H} \sim 100~{\rm cm^{-3}}~\left(\frac{T}{10^{7.5}~{\rm K}}\right)^{3/2}\left(\frac{m_{\rm SPH}}{10^4~\Msun} \right)^{-1/2}. 
\end{equation}
As in \citet{Schaye15}, we introduce a multiplication factor $f_{\rm th}$ to avoid the over-cooling problem even at the high-density star-forming region ($n_{\rm H}>100~{\rm cm^{-3}}$), 
\begin{equation}
    f_{\rm th} = f_{\rm th,min} + 
    \frac{f_{\rm th,max}-f_{\rm th,min}}{1+\left(\frac{Z}{0.1~\Zsun}\right)^{n_{\rm z}}\left(\frac{n_{\rm H,birth}}{ n_{\rm H,0}} \right)^{-n_{\rm n}} }, 
\end{equation}
where $n_{\rm H,birth}$ is the gas density where a stellar particle forms, $n_{\rm z}$, $n_{\rm n}$ and $n_{\rm H,0}$ are free parameters. 
The SN feedback is likely to be strong in low metallicity and low-density regions because of inefficient radiative cooling. 
Therefore, the multiplication factor is set as the function of metallicity $Z$ and the density $n_{\rm H}$ as  $n_{\rm z}=n_{\rm n}=2/\ln(10)$ and $n_{\rm H,0} = 0.67~{\rm cm^{-3}}$ \citep{Schaye15}. 
We set $f_{\rm th,max} = 2.5$ and $f_{\rm th,min}=0.3$ as the upper/lower limit of $f_{\rm th}$. 
The multiplication factor can exceed unity.  
This considers the additional feedback processes that are not included in the simulation, e.g., stellar winds, cosmic rays, and/or more energetic SN yields than assumed in the simulation. 
Note that the upper limit of the multiplication factor is lower than that of previous works \citep[$f_{\rm th, max}=3.0$ is chosen in][]{Schaye15} by considering the additional radiative feedback channels as described above. With the parameters, our previous work reproduced properties of observed high-redshift galaxies \citep{Yajima21}.
In our simulations, stellar particles continuously release hydrogen, helium, and metals into the neighbour gas. Based on the prescription in \citet{Wiersma09b}, we consider the gas and metal productions with tabulated yields for Types Ia and II SNe, and from asymptotic giant branch stars.

\subsection{Black holes}\label{sec:model_BH}
The AGN feedback is expected to suppress the star formation in massive galaxies via the heating and evacuating ISM \citep[e.g.,][]{Dubois12, Schaye15}.
Our simulations include the growth and the feedback processes of massive black holes (BHs) in galaxies. 
Since resolving an accretion disc around a BH is still difficult even in the current galaxy formation simulations,
we introduce subgrid models of the growth and feedback of BHs. 
First, a seed BH with the mass of $10^5~h^{-1}~\Msun$ is put at the galactic centre when the halo mass reaches $10^{10}~h^{-1}~\Msun$. 
As for the orbital evolution of BH particles, the simple repositioning technique is adopted to mimic the effect of dynamical friction on the BH. 
Recent work suggested that this technique is important to consider the feedback from supermassive BHs (SMBH) in the large-scale cosmological hydrodynamic simulations \citep{Bahe22}. 
We evaluate the gas accretion rate onto the BH particle based on the Bondi-Hoyle-Lyttleton model as
\begin{equation}
    \dot{m}_{\rm Bondi} = 
    \frac{4\pi c G M_{\rm BH}^2\rho}{\left(c_{\rm s}^2 + v_{\rm rel}^2\right)^{3/2}}, 
\end{equation}
where $\rho$, $c_{\rm s}$ are the density and the sound speed of gas particle around a BH, $v_{\rm rel}$ is the relative velocity between the gas and the BH. 
We consider the suppression of the gas accretion due to the angular momentum of the gas as
\begin{equation}
    \dot{m}_{\rm acc} = 
    \dot{m}_{\rm Bondi}\times {\rm min}(C^{-1}_{\rm visc} (c_{\rm s}/V_{\rm \phi})^3,~1), 
\end{equation}
where $C_{\rm visc}$ is a numerical parameter regarding the viscosity of the accretion disc, $C_{\rm visc}=200\pi$ is chosen as the fiducial setup \citep{Schaye15}. 
Using the accretion rate, the growth rate of BHs is estimated by $\dot{m}_{\rm BH} = (1-f_{\rm r})\dot{m}_{\rm acc}$, where $f_{\rm r} = 0.1$ is the radiative efficiency. 
The physical properties of neighbour gas are evaluated by using the nearest 100 gas particles. 
The fiducial model takes into account the upper limit of the accretion rate by the Eddington limit, 
\begin{equation}
    \dot{m}_{\rm Edd} = 
    \frac{4\pi G M_{\rm BH}m_{\rm p}}{f_{\rm r}\sigma_{\rm T}c}. 
\end{equation}
The FOREVER22 project  also includes the run with the optional  allowing the super-Eddington accretion, i.e., $\dot{m}_{\rm acc} > \dot{m}_{\rm Edd}$. 
Throughout this paper, we investigate only the results from  the fiducial models with the Eddington limit. 

In the simulations, during the timestep of $\Delta t$ the released energy is estimated as $\Delta E = f_{\rm e}f_{\rm r}\dot{m}_{\rm BH}c^2\Delta t$, where $f_{\rm e} = 0.15$ is the thermal coupling factor. 
As described below, the FOREVER22 project considers two types of feedback mechanisms, thermal (quasar mode) or kinetic (radio mode) feedback. 

\subsubsection{Quasar mode feedback}
In the quasar mode feedback, the released energy $\Delta E$ is injected into neighbour gas particles by heating the temperature up to $\Delta T = 10^9~{\rm K}$. 
Unlike the SN feedback, the gas-particle heated is selected from the nearest one.
If the released energy from a BH is large, the next particles are heated up in order of the distances.  
If the BH mass is smaller than $10^{9}~\Msun$, the quasar mode feedback alone is always used. 

\subsubsection{Radio mode feedback}
The jet-like kinetic feedback is activated when the BH mass exceeds $10^9~\Msun$. 
In radio mode feedback, half of $\Delta E$ is injected as the kinetic energy, and the other one is used as the thermal energy. 
The target gas particle of the radio mode feedback receives the momentum kick with the velocity of $3000~\rm km \; s^{-1}$. 
The direction axis of momentum is determined along the total angular momentum vector $\bm n = \bm L/|\bm L|$ of the neighbour gas particles. 
Then the direction $\bm n$ or $-\bm n$ is randomly chosen. 

\subsection{The Models}

The FOREVER22 project focuses on massive 10 PC regions  chosen in the (714.3~ cMpc)$^3$ simulation volume. 
The PC regions are explored with the different three sizes of zoom-in simulation volumes with different resolutions to cover the wide dynamic ranges; {\it Proto Cluster Region (PCR) run} (28.6~cMpc)$^3$, {\it Brightest proto-Cluster Galaxy (BCG) run} ($\sim$ 10~cMpc)$^3$ 
and {\it First run} ($\sim$ 3~cMpc)$^3$.
In this study, we refer to the results of PCR runs. 
The PCR runs allow us to investigate the large-scale gas structure around galaxies covering a wide range of their mass.
The PCR runs resolve the zoom-in volume with the gas and DM particle mass resolutions of $m_{\rm SPH}=2.9\times10^{6}~h^{-1}~\Msun$ and $m_{\rm DM}=1.6\times10^7~h^{-1}~\Msun$. The gravitational softening length is set to $2.0~h^{-1}~{\rm kpc}$. 
10 PC regions are labeled as PCR0 to PCR9 and the simulations are performed from $z=100$ to $2$. 
Most massive haloes in each PC region reach $\sim 10^{14}~\Msun$ at $z=2$. 
We have total 200 snapshots for each run, which is corresponding to the time bin $\Delta t\sim 16~{\rm Myr}$. 
We perform the post-processing analysis for the snapshot data to explore the gas dynamics and the statistical properties of very massive galaxies in detail.


\begin{figure*}
	\begin{center}
		\includegraphics[width=18cm]{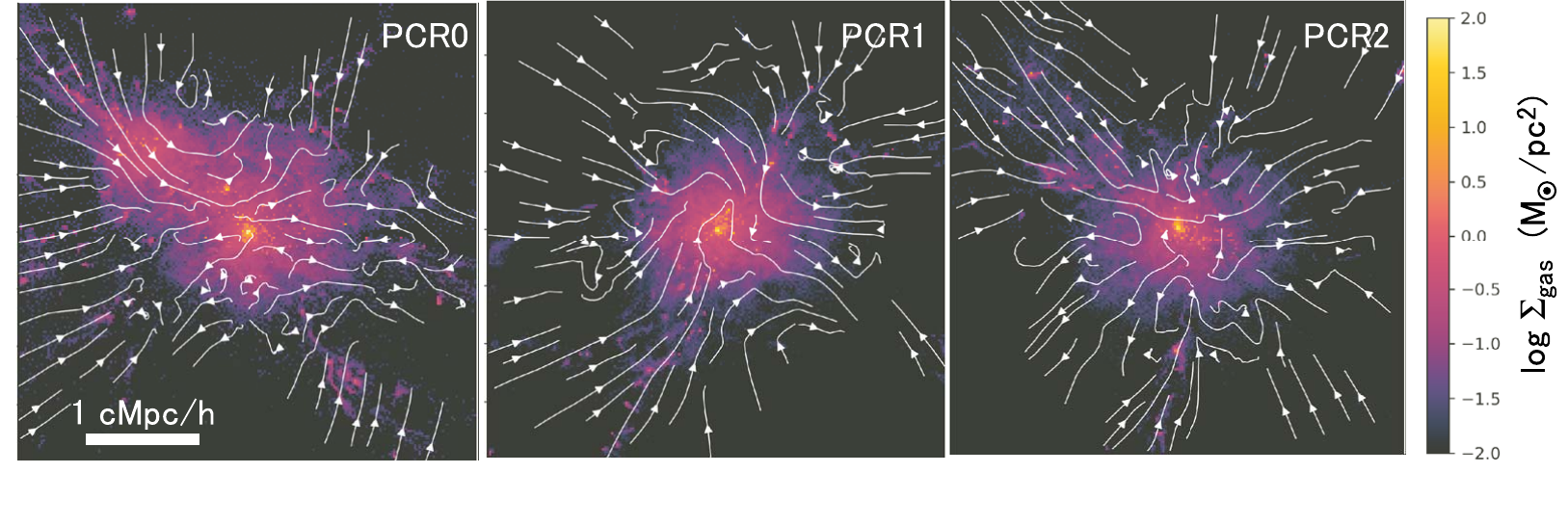}
	\end{center}
	\caption{Spatial distributions of gas column densities in most massive haloes at $z=3.0$ in the three protocluster
                 regions: PCR0, PCR1, and PCR2. The halo masses are $1.2 \times 10^{14}~\rm \Msun$ (PCR0), 
                 $8.4 \times 10^{13}~\rm \Msun$ (PCR1), and $5.5 \times 10^{13}~\rm \Msun/h$(PCR2). 
                 The projection depth is 50 $\rm ckpc~h^{-1}$ and white arrows represent
                streamlines of gas.
                 }
	\label{fig:map}
\end{figure*}


%
%

\section{GAS INFLOW AND OUTFLOW IN MASSIVE HALOES}
\subsection{Halo mass dependence of outflow properties}\label{sec:3.1}

Fig.~\ref{fig:map} shows structures of gas flow around most massive haloes at $z=3$ in PCR0, PCR1, and PCR2 runs. There are large-scale inflow structures along IGM filaments. Inflow and outflow coexist near star-forming regions, and some streamlines show fountain structures. 
Below, we statistically evaluate the inflow and outflow properties. 

We evaluate the inflow and outflow rates as in \citet{Nelson19}.
First, we estimate the radial velocity as
%
$
v_{\rm r,i} = \frac{\boldsymbol {v}_{\rm i}(t) \cdot \boldsymbol {r}_{\rm i}}{|{\boldsymbol {r}_{\rm i}}|}
$
%
,  where $\boldsymbol {v}_{\rm i}$ is the velocity vector of $i$-th gas particle, and define the outflow (inflow) if it is positive (negative). The position vector of $i$-th particle $\boldsymbol {r}_{\rm i}$ is estimated from the centre of mass of stellar component.
Then, we consider all outflowing gas particles of radial distances in the range between $r$ and $r+\delta r$ and estimate the outflow rate as
\begin{equation}
\label{flow}
\dot{M}_{\rm out}(r) = \sum_{i=1}^n \frac {m_{\rm i}}{\delta r}|v_{\rm r,i}|,
\end{equation}
where $m_{\rm i}$ is mass of $i$-th gas particle. 
In the case of inflow, we consider only gas particles with negative radial velocities. Note that, here the inflow gas includes accretion from IGM, galaxy mergers, and returning ones after evacuation from star-forming regions as the outflow.

\begin{figure*}
	\begin{center}
		\includegraphics[width=14cm]{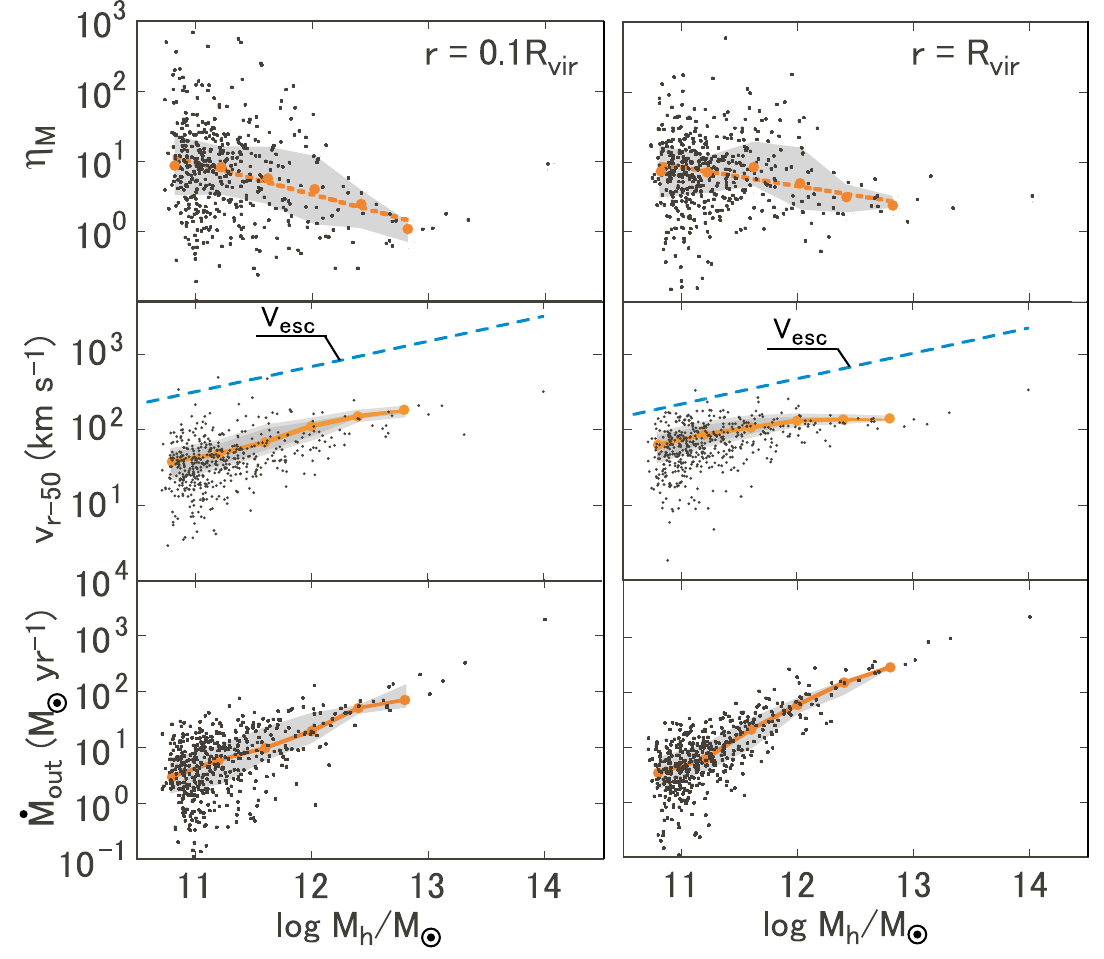}
	\end{center}
	\caption{		
		 Halo mass $\Mh$ scaling of the mass loading factor $\eta_{\rm M}$, median of the radial 
                 velocity $v_{\rm r-50}$, and the outflow rate 
                 $\dot{M}_{\rm out}$ extracted from top 500 massive haloes in PC0 at $z = 3.16$.  Red circles indicate the 
                 median and gray shades show the quantiles (25 - 75 percent) in each bin with a bin width of 0.4 dex. Red dashed lines 
                 in top panels are fitting results of medians by the power law, 
                 $\eta_{\rm M}=7.43 \times 10^{5}(\Mh/\Msun)^{-0.45}$ for the left top panel and 
                 $\eta_{\rm M}=7.24 \times 10^{3}(\Mh/\Msun)^{-0.27}$ for the right top panel. Blue lines in the middle panels are the escape velocity estimated assuming the 
                 NFW profile. In the left panels, each quantity is evaluated at radius $r = 0.1\Rvir$ as galaxy-scale 
                 outflow whereas at $r = \Rvir$ as halo-scale outflow in right panels. 
                }
	\label{fig:MLF-Mh}
\end{figure*}

In this work, we measure the inflow and outflow rates at $r=0.1$ or $1.0 \; R_{\rm vir}$. The former ones indicate “galaxy-scale” outflow, whereas the latter ones do “halo-scale” outflow. 
Fig.~\ref{fig:MLF-Mh} presents the outflow rate $\dot{M}_{\rm out}$,  
mass loading factor $\eta_{\rm M}$, and outflow velocity at $z=3$. 
 To understand the relationship between the feedback and the halo gravitational potential, we show the halo mass scaling of these values rather than the stellar mass. 
Here we consider the top 500 massive haloes identified by the standard friend-of-friend (FoF) grouping technique (containing central and satellite galaxies), corresponding to the mass range from $6.3 \times 10^{10}~\Msun$ to $1.0 \times 10^{14}~\Msun$. 
The mass loading factor is evaluated as $\eta_{\rm M} = \dot{M}_{\rm out}/{\rm SFR}$.  
We find that, at $r = 0.1\Rvir$, $\eta_{\rm M}$ monotonically decreases with increasing halo mass at the mass range of $\Mh \lesssim 10^{13}~\Msun$, showing $\eta_{\rm M} \sim 8.8$ at $\Mh = 10^{10.8}~\Msun$ and $\eta_{\rm M} \sim 1.1$ at 
$\Mh = 10^{12.8}~\Msun$. On the other hand, the trend of $\eta_{\rm M}$ 
reverses at the halo mass higher than $10^{13}~\Msun$ and the most massive halo shows $\eta_{\rm M}=9.3$.
Such a trend is also seen in some previous studies. \citet{Nelson19} found an upturn at the stellar mass of $\sim 10^{10.5}~\Msun$ in Illustris TNG50 simulation at $z = 2$, while \citet{Mitchell20} showed it occurred at $\Mh \sim 10^{12}~\Msun$ in EAGLE simulation at $z < 7.3$. They suggested the upturn was due to the strong BH feedback in massive haloes. 
Reasonable AGN feedback models are still under debate and depend on the resolution and simulation schemes. The feedback efficiency $f_{\rm e}$ and the energy fraction deposited into the kinetic or thermal form are quite different among different simulations. Our model is basically similar to that in EAGLE \citep{Schaye15} but includes the jet-type feedback as in illustrisTNG \citep{Nelson19}. 
Note that, however, the number of such massive haloes is not sufficient for statistically robust discussion due to random variability within the cosmological simulations \citep[]{genel19, Keller19}.

Using the median values at $\Mh < 10^{13}~\Msun$, we derive a fitting with a form, ${\rm log_{10}}\eta_{\rm M} = \alpha {\rm log_{10}} \Mh/{\Msun} + \beta$. We find the best-fit parameters as $\alpha=-0.45$ and $\beta=5.9$. This slope is intermediate between two characteristic  scaling law in $\eta_{\rm M}-\Mh$ relation, one is the momentum-conservation model ($\eta_{\rm M} \propto \Mh^{-1/3}$) and another is the energy-conservation one ($\eta_{\rm M} \propto \Mh^{-2/3}$) \citep[]{Murray05,  Veilleux20}. The reasonable model sensitively depends on the cooling efficiency at a post-shock region. The adiabatic case is corresponding to the energy-conservation model. Our results suggest that the cooling is moderately effective on the galactic scale. 

In the halo scale,  $\eta_{\rm M}$ slowly decreases at $\Mh \lesssim 10^{13}~\Msun$ and becomes almost flat at $\Mh \gtrsim 10^{13}~\Msun$. 
We perform the fitting at $M_{\rm h}<10^{13}~\Msun$ as with the galaxy-scale analysis, and the best-fit slope is $\alpha = -0.27$. 
Therefore, we indicate that the outflow propagates with the momentum-conservation mode in the halo scale. 
 Our results at $\Mh \lesssim 10^{13}~\Msun$ are similar to \citet{Mitchell20} in the galaxy scale, but the slope is shallower than their work ($\alpha \sim -1$) in the halo scale. This may indicate that in our simulations the cooling works efficiently rather than \citet{Mitchell20} at the halo scale. 
Although the number of samples is not sufficient to discuss the statistical features, our simulation contains the more massive haloes. 
Our result shows
 even massive haloes with $\Mh \sim 10^{14}~\Msun$ can keep $\eta_{\rm M}$ greater than unity that indicates protoclusters in the early Universe can provide gas and metals into IGM.

Radial velocities of outflow gas are also shown in the figure.
In the galactic-scale, median values of the outflow gas ($v_{\rm r-50}$) monotonically increase with the halo mass from $38~\rm km\; s^{-1}$ at $\Mh=10^{10.8}~\Msun$ to $304 ~\rm km\; s^{-1}$ at $10^{14}~\Msun$.
One of the characteristic velocities of the galactic halo is the escape velocity $V_{\rm esc} = (2G\Mh/\Rvir)^{1/2}$ which is equal to $\sqrt{2}$ times the circular velocity $V_{\rm c}$. 
Assuming $\Mh \propto \Rvir^{3}$, we obtain the halo mass scaling of escape/circular velocities as $\propto \Mh^{1/3}$. 
In the middle panel of Fig. \ref{fig:MLF-Mh}, we overplot $V_{\rm esc}$ as the function of $M_{\rm h}$. 
Our simulations show that $v_{\rm r-50}$ has a halo mass dependence similar to  $V_{\rm esc}$ whereas the absolute value is smaller. 
\citet{Martin05} have pointed out the tight relationship between the outflow velocity and the circular velocity. 
The velocity ratio of $v_{\rm r-50}$ to $V_{\rm c}$ is $\sim 0.4$ at $10^{10.8}~\Msun < \Mh < 10^{12.8}~\Msun$. 
For halo-scale outflow, however, the increasing trend 
disappears beyond $\Mh \sim 10^{12}~\Msun$.
The difference between $V_{\rm esc}$ and $v_{\rm r-50}$ increases with the halo mass, suggesting that most gas particles cannot escape from massive haloes. This will be discussed in  Section \ref{sec:Trajectories}  and 4.

The bottom panels show outflow rates. As shown in \citet{Yajima21}, massive galaxies are on the observed main-sequence \citep{Pearson18}, having SFR of $\gtrsim 1000~\Msunyr$. A similar amount of gas is evacuated, resulting in $\eta \gtrsim 1$.
Given that the gas accretion rate onto a halo is low, most gas can be lost within the cosmic time scale. Note that, however, \citet{Yajima21} suggested that most massive haloes in PC regions keep gas-rich states. 

\begin{figure*}
	\begin{center}
		\includegraphics[width=15cm]{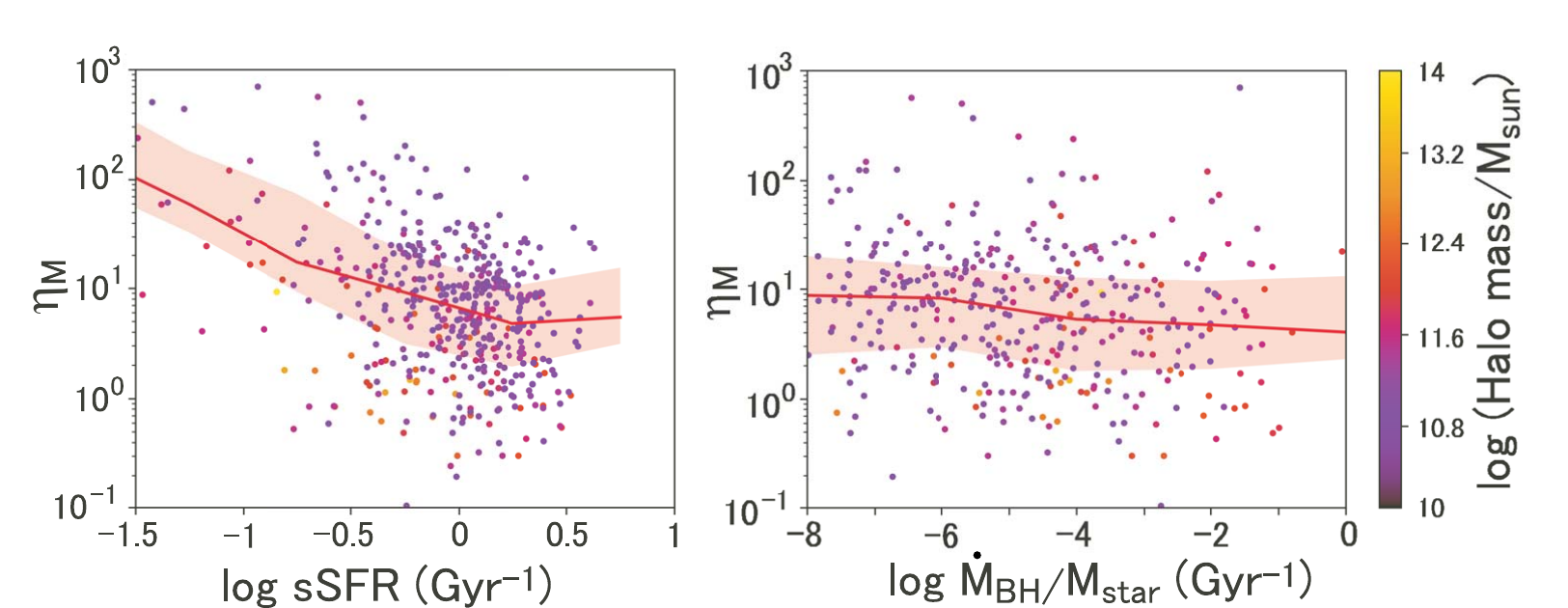}
	\end{center}
	\caption{		
		 left panel: 
		 Mass loading factor $\eta_{\rm M}$ as a function of sSFR and gas accretion rate onto BHs divided by stellar mass at $r=0.1~R_{\rm vir}$. The top 500 massive haloes at $z=3.16$ are analyzed. Red lines and shades show the median values and quartiles (25 - 75 percent) in each bin. Each point is classified by color which indicates the  halo mass.
                }
	\label{fig:MLF-sSFR}
\end{figure*}

Massive stars and BHs are thought to be the two main drivers of the galactic outflow.
Fig.~\ref{fig:MLF-sSFR} shows the dependence of $\eta_{\rm M}$ on specific star formation rate (sSFR) and gas accretion rate onto BH for $r=0.1~R_{\rm vir}$. 
We can find that massive haloes have a lower mass loading factor as compared with low-mass haloes even if they have similar sSFR or $\dot{M}_{\rm BH}$. This is consistent with the negative $\eta_{\rm M}-M_{\rm h}$ relation in Fig. 2. 
We find that $\eta_{\rm M}$ decreases with sSFR. 
Low-mass haloes show frequent fluctuations in star formation histories due to the cycles of outflows and refreshing of gas \citep[see also,][]{Yajima17c, Arata20, Abe21}. Then, galaxies can keep high star formation rates as the halo mass increases. Therefore, the low values of sSFR may indicate the phase of the suppressed star formation after a starburst. 
 Note that the fluctuations of SFR can also be shown in the case with a merging resolution following the structure low-mass galaxies \citep[e.g.,][]{Crain15}.
Considering the time lag between the launching due to the starburst and the large-scale outflow  $0.1 R_{\rm vir} / V_{\rm c} \sim  31.5\left( \frac{1+z}{4} \right)^{-3/2}~{\rm Myr}$, the anti-correlation can be explained. 
The decreasing tendency for $\eta_{\rm M}$ shifts to flat ( orange line) or slightly increasing tendency (shade) at ${\rm sSFR}\gtrsim 1~\rm Gyr^{-1}$. 
A part of galaxies with ${\rm sSFR} \gtrsim 1~\rm Gyr^{-1}$ are thought to be massive galaxies in the starburst phase. 
As the result of a large amount of SN energy from stars, the mass loading factor may keep the constant value slightly less than 10 (as the median value). 
Contrary to the result for sSFR, we find no strong dependency of $\eta_{\rm M}$ regarding the BH activities. 
 Here, the BH activity is measured at each time step. However, it can change rapidly over time scales shorter than the dynamical time of the galactic outflow. The time-scale difference may weaken the dependence. 
Note that the energetic AGN feedback has the potential to drive the outflowing gas comparable to or greater than that of stars. 
As we described in \S\ref{sec:model_BH}, the BH particles release the energy of $\Delta E = f_{\rm e}f_{\rm r}\dot{m}_{\rm BH}c^2\Delta t$ during the time interval of $\Delta t$. 
Thus, we estimate the releasing energy rate of a BH as
\begin{eqnarray}
    \dot{E}_{\rm BH} &=& f_{\rm e}f_{\rm r}\dot{m}_{\rm BH}c^2 \nonumber \\
    &=& 8.5 \times 10^{44}~{\rm erg~s^{-1}}~
    \left( \frac{f_{\rm e}f_{\rm r}}{0.015} \right) \left( \frac{\dot{m}_{\rm BH}}{\Msunyr}\right)
\end{eqnarray}
On the other hand, the available releasing SN energy per unit mass is $\epsilon_{\rm SNII}=8.73\times10^{15}~{\rm erg~ g^{-1}}$ by assuming the Chabrier IMF and the energy from a single event of SN as $10^{51}~{\rm erg}$ \citep{DallaVecchia12}. 
Considering the continuous star formation over $\sim 10$~Myr (typical lifetime of the massive stars), 
the releasing energy rate due to the SN feedback may become as 
\begin{equation}
    \dot{E}_{\rm SN} = \epsilon_{\rm SNII}\times{\rm SFR}
    =5.5\times10^{43}~{\rm erg~s^{-1}}~
    \left(\frac{{\rm SFR}}{100~\Msunyr}\right). 
\end{equation}
In the case of the energy conservation outflow, the releasing energy rate $\dot{E}$ is equated with the production rate of kinetic energy as $ \frac{1}{2}\dot{M}_{\rm out}V_{\rm out}^2 = \dot{E}$. 
On the other hand, in the case of the momentum conservation mode, the production rate of the momentum  can be estimated as $\dot{M}_{\rm out} V_{\rm out} = \dot{P} \sim \dot{N}m_{\rm SPH}v_{\rm th} \sim \dot{E}/v_{\rm th}$, where $v_{\rm th}$ is the thermal velocity of heated gas particles due to the SN/BH feedback. 
Given that $V_{\rm out} \sim V_{\rm esc}$, the ratio of the outflow rates due to the BH and the SN feedback  is evaluated as
\begin{equation}
    \frac{\dot{M}_{\rm out}^{\rm BH}}{\dot{M}_{\rm out}^{\rm SN}}  
    =\xi \left(\frac{f_{\rm e}f_{\rm r}}{0.015}\right) \left(\frac{\dot{m}_{\rm BH}/{\rm SFR}}{0.01}\right),  
\end{equation}
where the factor $\xi = 15.3$ 
for the energy conservation case and $\xi = 2.7 \left(\frac{10^{7.5}~{\rm K}}{\Delta T_{\rm SN}}\right)^{1/2}\left(\frac{10^{9}~{\rm K}}{\Delta T_{\rm BH}}\right)^{-1/2}$ for the momentum conservation case, respectively.  
Hence, the BH feedback would contribute to the outflow rate if $\dot{m}_{\rm BH}/{\rm SFR}$ is larger than $6.5\times 10^{-4}~ (3.7 \times 10^{-3})$ for the energy (momentum) conservation case. As indicated in \citet{McAlpine17}, BHs can grow rapidly once the halo mass exceeds $\sim 10^{12}~\rm M_{\odot}$. 
A similar trend is shown in FOREVER22 simulation \citep[see Fig. 8 and 11 in][]{Yajima21}.  
In the numerical simulations, 
the activity of BH feedback drastically changes over time 
and we find no clear dependency between $\eta_{\rm M}$ and $\dot{M}_{\rm BH}$.  The BH growth and activity are quite different among simulations \citep[e.g.,][]{Dave19, Dubois21}. However, \citet{Nelson19} showed a similar weak dependency regardless of the different resolution and BH models.
However, the BH feedback has the potential to boost the $\eta_{\rm M}$ even in the high-mass galaxies with the mass $M_{\rm h}\gtrsim 10^{12}~\Msun$ (see Fig. \ref{fig:MLF-Mh}). 

\subsection{Radial profile of inflow and outflow rate}

\begin{figure*}
	\begin{center}
		\includegraphics[width=15cm]{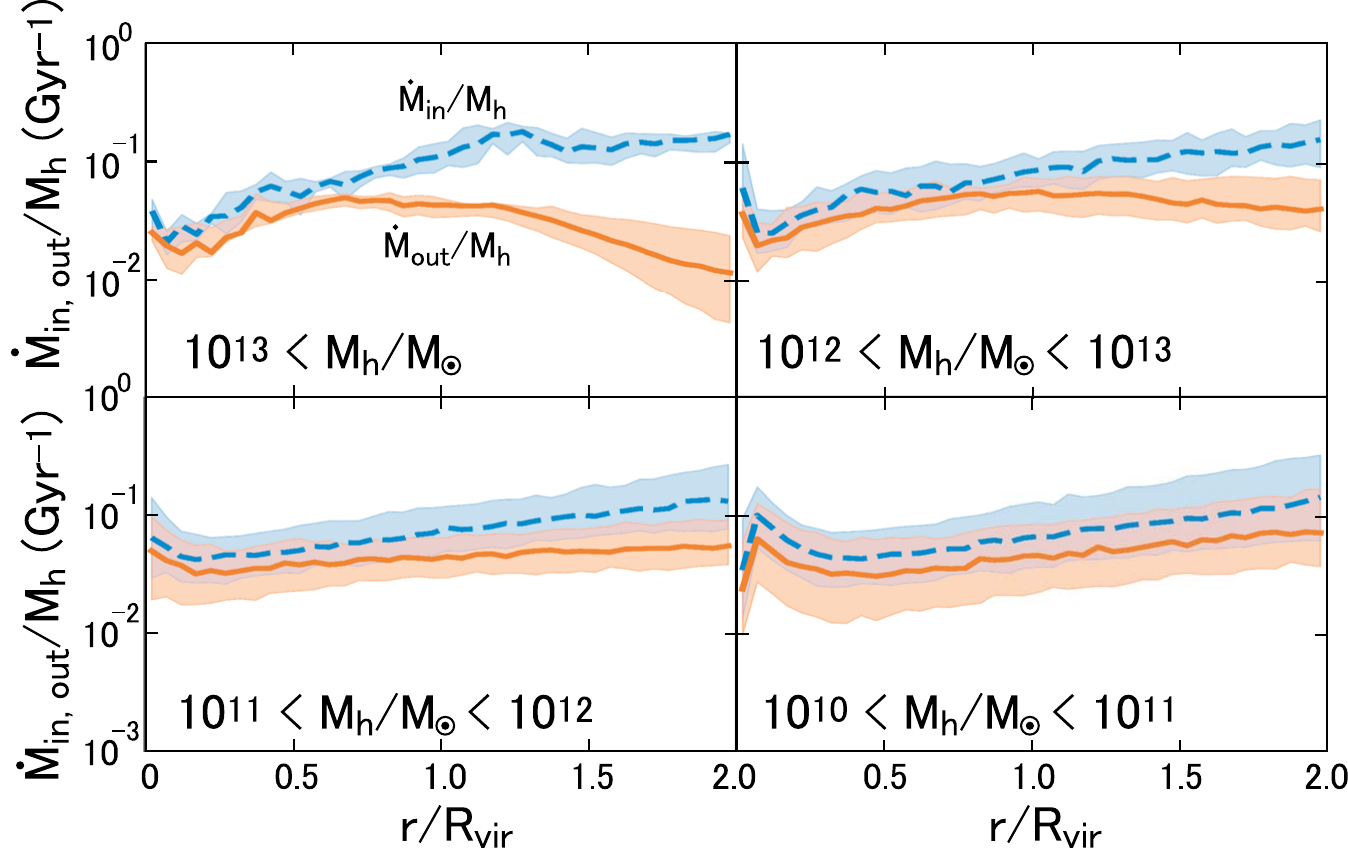}
	\end{center}
	\caption{
	Radial profiles of inflow and outflow rates divided by halo masses at $z=3.0$. The inflow/outflow rates are evaluated at $r=\Rvir$. 
 Red solid and blue dashed lines are the median values of outflow and inflow rates in each halo mass range. Shades represent the quartiles (25-75 percent).
                }
	\label{fig:flow-r}
\end{figure*}

 The velocity of the outflow gas gradually decreases as interstellar matter or CGM is stacked. Once the velocity reaches the escape velocity at a specific radius, the outflow gas starts to return as the inflow. To study the point,
we examine radial profiles of inflow and outflow rates. Fig.~\ref{fig:flow-r} presents the specific inflow and outflow rates defined as $f_{\rm s-in,s-out} \equiv \dot{M}_{\rm in, out}(r) /\Mh$. We divide haloes into four mass bins and measure the median values.   
$f_{\rm s-out}$ of the most massive group (top-left-hand panel) declines at $r \gtrsim R_{\rm vir}$.  This indicates a part of gas returns into a halo at $r \sim R_{\rm vir}$ as a halo-scale fountain flow. As the radial distance increases, the difference between 
$f_{\rm s-in}$ and $f_{\rm s-out}$ increases and reaches around ten times.
In cases of lower mass haloes, the outflow rates keep constant or increase even beyond $r \sim R_{\rm vir}$, suggesting that metal-enriched gas from star-forming regions is provided into IGM. The specific outflow rates exceed $\sim 3 \times 10^{-2}~\rm Gyr^{-1}$.
Therefore, the low-mass haloes can lose most gas within the cosmic age at $z \sim 3$ considering the gas fraction $M_{\rm gas}/\Mh \sim \Omega_{\rm b}/\Omega_{\rm M}$ if no gas inflow exists. However, we confirm that $f_{\rm s-in}$ is higher than $f_{\rm s-out}$ overall radius within $2 R_{\rm vir}$. Therefore, most high-redshift galaxies grow in a gas-rich state.
 \citet{Mitchell20} also showed the radial profiles of the outflow gas within a viral radius at $z=0$. Our simulation results 
 show similar structures that the outflow rate increases with the radial distance within a virial radius. Note that, the outflow rate is higher than the figure of the radial profile in \citet{Mitchell20} by a factor of few because of the different redshifts. Here, by expanding the range of the radial distance and the halo mass, we show that the rates of the outflow gas rapidly decrease just beyond the virial radii in the case of massive haloes.

 The structures of the inflow and the outflow gas are likely to be reflected in properties of absorption lines in SEDs of background objects like QSOs or bright galaxies, which will be one of the main topics carried out by future observational missions, e.g., the Subaru Prime Focus Spectrograph. \citet{Wright20} suggested that the baryon accretion rate sensitively depended on the sub-grid physics like stellar or AGN feedback. In particular, the accretion rate for low-mass and low-redshift galaxies is sensitive to the sub-grid models because of the efficient galaxy outflows. We will investigate the reasonable sub-grid models via comparisons with upcoming observational data in future work.

\subsection{Redshift evolution of inflow and outflow}
\begin{figure*}
	\begin{center}
		\includegraphics[width=18cm]{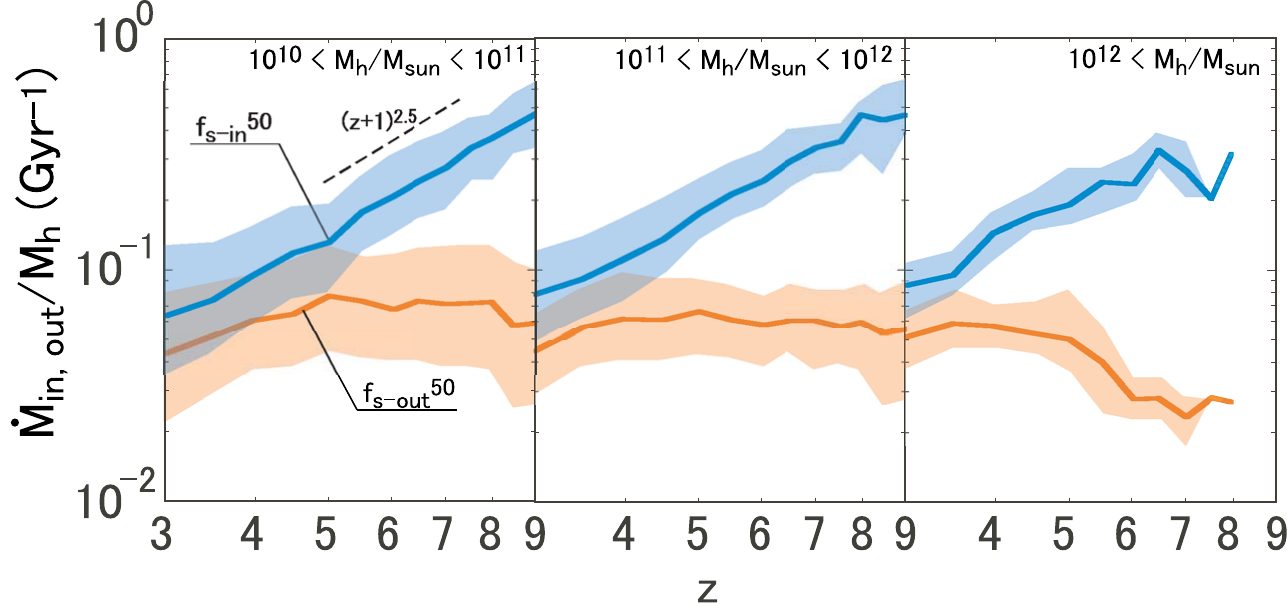}
	\end{center}
	\caption{
                 Redshift evolution of outflow and inflow rates normalized by halo masses for three halo groups:  $10^{10} < \Mh/\Msun < 10^{11}$ (left panel),  $10^{11} < \Mh/\Msun < 10^{12}$ (center panel), and  $10^{12} < \Mh/\Msun$ (right panel). Red and blue lines are the median values of outflow and inflow rates in each halo group, and shades represent the quartiles (25-75 percent).
                }
	\label{fig:tdepend}
\end{figure*}

Fig.~\ref{fig:tdepend} shows the redshift evolution of the specific gas inflow/outflow rate at $r = \Rvir$. 
We pick up haloes with the stellar mass larger than $10^{8}~\Msun$ and divided them into three groups according to their halo mass $\Mh$ at each redshift.  
Overall, we find that there are no clear differences between the three groups. 
All cases indicate an increasing trend in $f_{\rm s-in}$ between $3 < z < 9$. 
\citet{Dekel13, Dekel14} developed a simple toy model on the galaxy evolution and showed the mass accretion rate as a function of the scale factor, $a=(1+z)^{-1}$. 
Given the Einstein-de Sitter universe, the specific mass accretion rate in the galaxy by the dark matter 
shows a monotonical increase with $z$ and is proportional to $(1+z)^n$, $n = 5/2$. \citet{Dekel13} 
assumed that the specific gas inflow rate is also expressed with the same relation and confirmed it by their simulations \citep{Dekel14}. 
In the cases with the halo masses of $< 10^{12}~\Msun$, the inflow rates increase with redshift as $(1+z)^{2.5}$, which indicates the haloes grow via mergers with gas-rich haloes simply. In the case of massive haloes with $\Mh > 10^{12}~\Msun$, the slope is somewhat shallower. This can be due to matter accretion or halo mergers with a low baryon fraction. 
The outflow rates, on the other hand, 
is remarkably different from the inflow and it does not show clear evolution. It keeps fairly 
constant with $z$ between $3 < z < 9$, whereas it shows some small variations, the increasing 
trend at $z \sim 3$, and the decreasing feature at high redshifts. Since the difference between 
the inflow and the outflow produce the change of baryon mass $M_{\rm b}$ in haloes, the large 
difference at high redshift suggests active growth in haloes. At a lower 
redshift of $z \sim 3$, however, the difference becomes small and the baryon mass growth may be suppressed. 
Note that, $f_{\rm s-out}$ of the massive halo group decreases at $z \gtrsim 6$. This is likely due to the deep gravitational potential that hosts most gas against the SN feedback. 

\subsection{Trajectories of outflowing gas particles}\label{sec:Trajectories}
\begin{figure}
	\begin{center}
		\includegraphics[width=8cm]{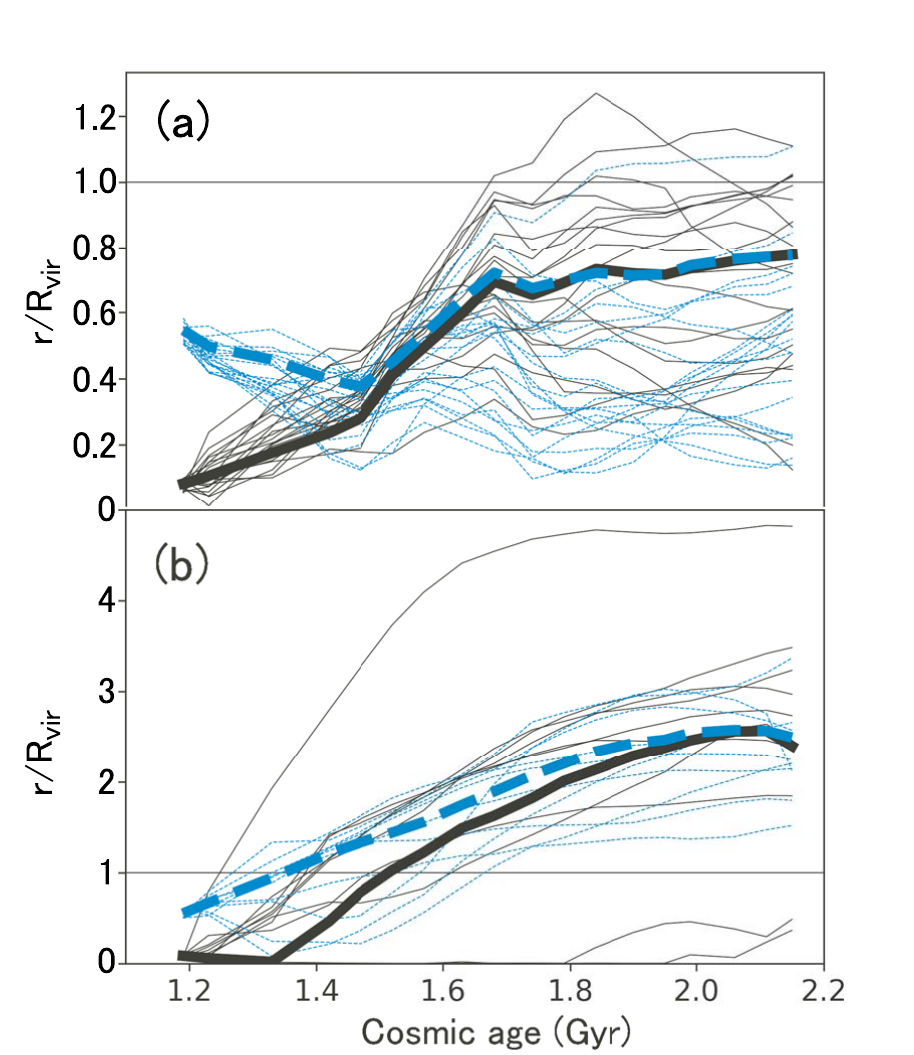}
	\end{center}
	\caption{		
		 Radial motion of outflowing gas particles in the most massive halo 
                 (panel (a), $\Mh \sim 1.3 \times 10^{14}~\Msun$ at $z = 2.89$) and  Halo37 
                 (panel (b), $\Mh \sim 1.1 \times 10^{12}~\Msun$ at $z = 2.89$) from $z = 4.79$ to $2.89$, 
                 corresponding to $0.96$ Gyr as the time interval.
                 At the initial step ($z = 4.79$), some outflow particles are categorized into two groups depending on their initial distances:  $0.05 < r/\Rvir < 0.1$ (black lines) and  $0.5 < r/\Rvir = 0.6$ (blue lines).
                 Thick lines show
                 median values of all outflow particles in each category. Thin lines represent the trajectories of  
                 10 particles in each group.
                }
	\label{fig:r-t}
\end{figure}

Thanks to the Lagrangean format of our code, we can track outflow gas particles \citep[]{Oppenheimer08, Ford14, Christensen16, Alcazar17, Borrow20, Hafen20}.
Fig.~\ref{fig:r-t} shows radial distances of the outflow gas particles of two sample haloes with the halo masses of $1.3 \times 10^{14}$ (Halo0) and $1.1 \times 10^{12}~\Msun$ (Halo37).
We follow the motions of gas particles from $z=4.8$ to $z=2.9$, corresponding to the time period of $1.0$ Gyr.
We classify target particles into two groups to study galaxy scale and halo scale outflow. 
For the galaxy-scale samples (group A), all gas particles with initial radial positions between $0.05$ and $0.1\Rvir$ are selected. For the halo scale (group B), we consider gas particles between $0.5$ and $0.6~\Rvir$.
The black and blue thick lines respectively represent  
median values for groups A and B at each time, while thin lines show trajectories of randomly selected 10 particles. 
As shown in the top panel of Fig. \ref{fig:r-t}, in Halo0,  both median values reach $r \sim 0.8 ~\Rvir$ at $t \sim 1.8~\rm Gyr$ and keep staying a similar distance. This indicates that most outflow gas contributes to hot halo gas. 
We find that a part of gas particles return to a galactic centre after reaching the radius of $\gtrsim 0.5~\Rvir$, i.e., the large-scale fountain flows (see the black thin line in the top panel of Fig. \ref{fig:r-t}). 
This feature indicates that most outflow gas have outflow velocities similar to halo escape velocities but it does not exceed them. Thus, in the case of massive haloes, the outflow rate decreases near the virial radius as shown in Fig.~\ref{fig:flow-r}.
In Halo37 (bottom panel in Fig. \ref{fig:r-t}), on the other hand, most gas particles 
exceed the virial radius within $\sim 0.3$ Gyr and reach $\sim 2.5 ~\Rvir$ at $t \sim 1$ Gyr finally. Thus, we suggest that  the baryon cycle between the galaxy and CGM sensitively depends on the host halo mass, which will be investigated with future observations with the Prime Focus Spectrograph on the Subaru telescope.

\begin{figure}
	\begin{center}
		\includegraphics[width=8cm]{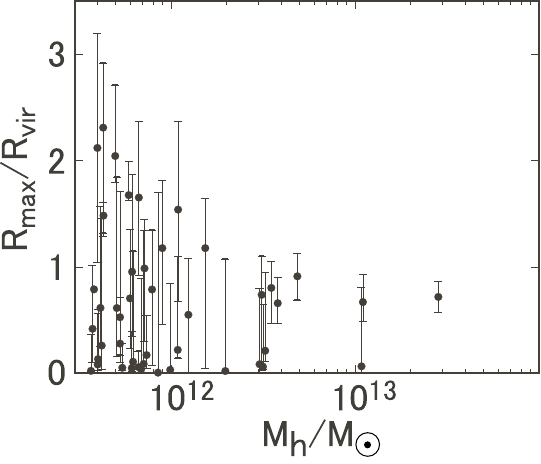}
	\end{center}
	\caption{
	        Maximum distances of outflow gas particles at $3.0 < z < 4.0$ for 44 massive haloes. The halo mass is measured at $z=4.0$. Only outflow particles with the initial positions of $0 < r/\Rvir < 0.1$ are chosen at $z=4.0$. Black-filled circles and error bars represent the median values of the outflow particles and the quartiles (25-75 percent).
	        }
	\label{fig:Rmax-Mh}
\end{figure}

We measure the maximum distances of outflow gas particles ($R_{\rm max}$). 
Fig.~\ref{fig:Rmax-Mh} shows the mass dependence of $R_{\rm max}$ by using 44 haloes with the mass range from 
$3.66\times 10^{11}~\Msun$ to $2.83\times 10^{13}~\Msun$ at $z = 3.0$. Here we pick the gas particles distributed within $0.1~\Rvir$ initially and consider their trajectories with the redshift range from $z = 4.0$ to $3.0$.
We confirm that these 44 haloes are always major progenitors of the haloes at $z=3$, i.e., they do not experience mergers with heavier haloes, because it is difficult to distinguish fountain flows and accretion onto merger companions. 
We find that most gas in haloes with a mass larger than $10^{12}~\Msun$ do not go beyond virial radii and $R_{\rm max}$ distributes at $\sim \Rvir$ or $ \lesssim 0.5~\Rvir$. On the other hand, some haloes of $\Mh < 10^{12}~\Msun$ show the large $R_{\rm max}$. Note that, in the case of haloes of $\Mh < 10^{12}~\Msun$, there is a large dispersion with the range between $R_{\rm max} \sim 0.02$ and $2.3$ that can be induced gas structures around star-forming regions. 
\citet{Borrow20} showed that a large fraction of gas can move beyond $1~\rm Mpc$ even in the case of massive galaxies by $z=0$. They suggested that the jet feedback could induce the outflow. Therefore, the migration distance of the gas can depend on the redshift and the feedback models.
If young stars are embedded in dense clouds even at the end of the lifetime of massive stars $\sim 10~\rm Myr$, thermal energy released by SN feedback can be lost via radiative cooling before the galactic wind is launched. 
 The haloes with the low values of $R_{\rm max}$ are likely to keep star formation smoothly, while the star formation of the ones with the large $R_{\rm max}$ can be quenched until the gas is recovered. The quenched time scale  can be $\tau \sim \Rvir/V_{\rm c} \sim 315~  \rm Myr$ at $z \sim 3$. 

\section{Metal outflow}
In this section, we investigate metal outflow from galaxies. 
The metal-enriched CGM can be a powerful tool to reveal the feedback and the star formation processes via  comparisons between observations and theoretical models \citep[e.g.,][]{Tumlinson17, Fujimoto19}. 
Also, the metal distribution in IGM can reflect the star formation history over the cosmic age \citep[e.g.,][]{Cowie98, Aguirre01}. 
 The complex nature of the metal recycling and the distribution would be related to not only the star formation but also the AGN activities \citep{Moll+07}. Previous numerical simulations investigated the outflow of metals from galaxies while considering the variation of halo mass. For instance, \citet{Christensen+18} performed high-resolution cosmological hydrodynamic simulations to investigate the history of metal recycling from dwarf to Milky Way-sized galaxies. They found that the gas outflow effectively removed metals from galactic discs and the metal was distributed dispersedly even beyond virial radii, while more massive galaxies can retain more metals within the virial radii. Recently, \citet{Pandya+21} performed the state-of-art cosmological hydrodynamic simulation and argued the properties of the gas/metal outflow. They found that the metal loading factor for low-mass galaxies could be an order of unity, which implies that most of the metals were ejected with the outflow. Also, for the low-mass galaxies, the ratio of metal loading factors for ISM and halo regions exhibited similar values, suggesting the efficient supply of the metals to CGM and IGM. However, the ratio drops for massive galaxies and the smaller fraction of the metals leaves the haloes. 
In this study, we extend the mass range of the galaxies up to $\gtrsim 10^{13}~\Msun$ and investigate the properties of metals between the galaxies and CGMs.  
 
\begin{figure*}
	\begin{center}
		\includegraphics[width=12cm]{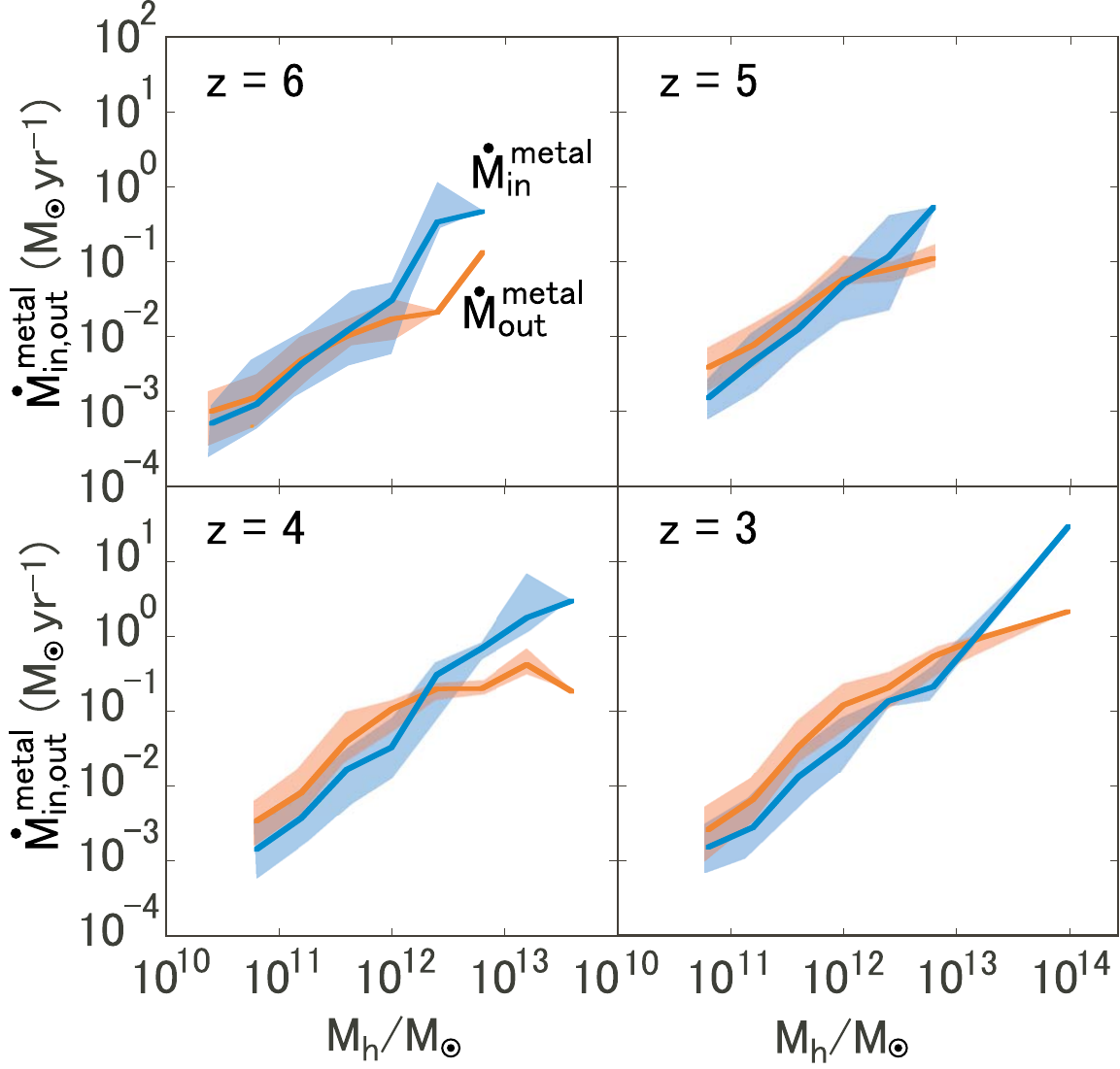}
	\end{center}
	\caption{	
	Metal outflow and inflow rates evaluated at virial radii. Red and blue  lines are the median values of outflow and inflow rates in each halo mass range. Shades represent the quartiles (25-75 percent).
                }
	\label{fig:metalflow}
\end{figure*}

Fig.~\ref{fig:metalflow} represents the  metal inflow and outflow rates at $r = \Rvir$ for 500 massive haloes at $z = 3, 4, 5$, and $6$. 
The inflow and outflow rates are calculated as
\begin{equation}
\dot{M}^{\rm metal}_{\rm in,out} = \sum_{i=1}^n \frac {m_{\rm i} Z_{\rm i}}{\delta r}|v_{\rm r,i}(r_{\rm i},t)|.
\end{equation}
Similar to the case of 
the gas flow, the metal outflow rate increases with the halo mass. At $z=3$, massive haloes induce the metal outflow larger than $1.0~ \Msunyr$, corresponding to the gas outflow of $\dot{M}_{\rm out} \gtrsim 1000~\Msunyr$ with the metallicity $Z \sim 0.1~\Zsun$.
 The mass loading factors for the massive galaxies show an order of unity at the virial radii (see Fig. \ref{fig:MLF-Mh}). The metal loading factor is evaluated as $\sim 10^{-3}$. 
Therefore, a small fraction of the metals are ejected into IGM in the case of massive galaxies. 

As expected from the results in the previous section, the metal outflow rate does not increase significantly at $M_{\rm h} > 10^{12}~\Msun$. On the other hand, the metal inflow rate simply increases with the halo mass via merger processes and accretion of metal-enriched gas around galaxies. In particular, the IGM around the massive haloes is metal-enriched earlier, which is likely due to the metal outflow from neighbour low-mass haloes. Therefore, the metal inflow rate becomes high at $\Mh \gtrsim 10^{12}~\Msun$. 
The metal outflow rate overcomes the inflow rate in the case of low-mass haloes and the trend is reversed at massive ones. Therefore, the process of metal enrichment within galaxies would be distinguishable depending on the halo mass. 
 For low-mass haloes, 
 the metal contents within the virial radius can be dominated by the metals produced in-situ star formation via supernovae.  
As the halo mass increases, 
the fraction of the metals supplied from outside the halo via accretion and mergers can be dominant. 
\citet{Christensen+18} investigated the galaxies with the mass range of $M_{\rm h}\sim 10^{9.5-12}~\Msun$ at $z = 0$ and pointed out the similar mass trend. 

\begin{figure}
	\begin{center}
		\includegraphics[width=8cm]{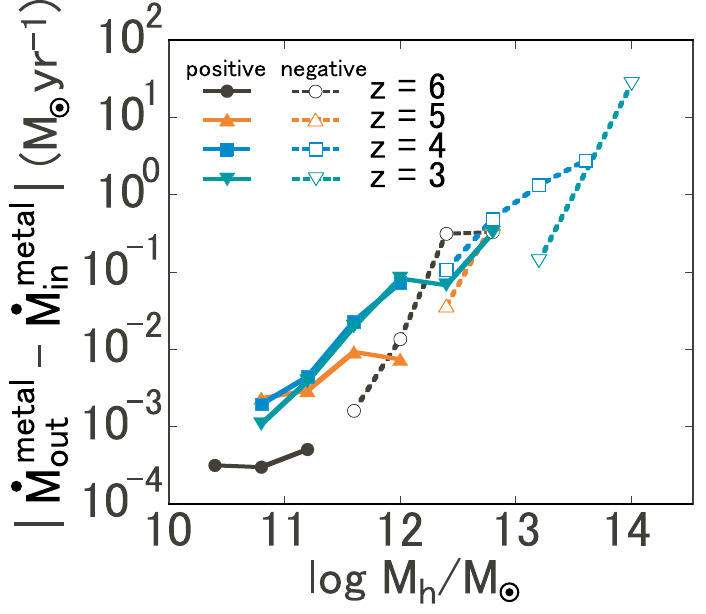}
	\end{center}
	\caption{	
	    Net outflow rate, $|\dot{M}^{\rm metal}_{\rm out}-\dot{M}^{\rm metal}_{\rm in}|$, where $\dot{M}^{\rm metal}_{\rm out}$ and $\dot{M}^{\rm metal}_{\rm in}$ are taken from Fig.~\ref{fig:metalflow}. Filled and open symbols show the cases of positive $\dot{M}^{\rm metal}_{\rm out}-\dot{M}^{\rm metal}_{\rm in}$ (outflow dominant) and negative ones (inflow dominant) respectively.
                }
	\label{fig:metalflow_difference}
\end{figure}

Fig.~\ref{fig:metalflow_difference} shows net metal outflow rates, i.e., $\dot{M}^{\rm metal}_{\rm out} - \dot{M}^{\rm metal}_{\rm in}$. We find that the net metal outflow changes from positive to negative as the halo mass increases.
The critical halo mass increases as the redshift decreases and it is $\sim 10^{13.0}~\Msun$ ($\sim 10^{11.2}~\Msun$) at $z = 3$ ($z=6$). 
 Because of deeper gravitation potentials at higher redshifts, the gas cannot escape from the haloes efficiently compared to the low-redshift haloes that have the same masses.
The moderately massive galaxies with $\Mh \sim 5 \times 10^{12}~\Msun$ have the net metal outflow rates of $\gtrsim 0.1~\Msunyr$.  
Also, the net-metal outflow rate of higher redshift is lower than that of lower redshift. This is likely due to lower metallicities of IGM and ISM at higher redshifts. Galaxies with the halo masses of $\gtrsim 10^{12}~\Msun$ show the metallicities 
 with $\sim 0.1-0.2~\Zsun$ at $z=6$ and it increases by a factor of $\sim 1.5-2.0$ at $z=3$.
Thus, low-mass haloes at lower redshifts can be efficient metal-enrichment sources of IGM. 

\begin{figure}
	\begin{center}
		\includegraphics[width=8cm]{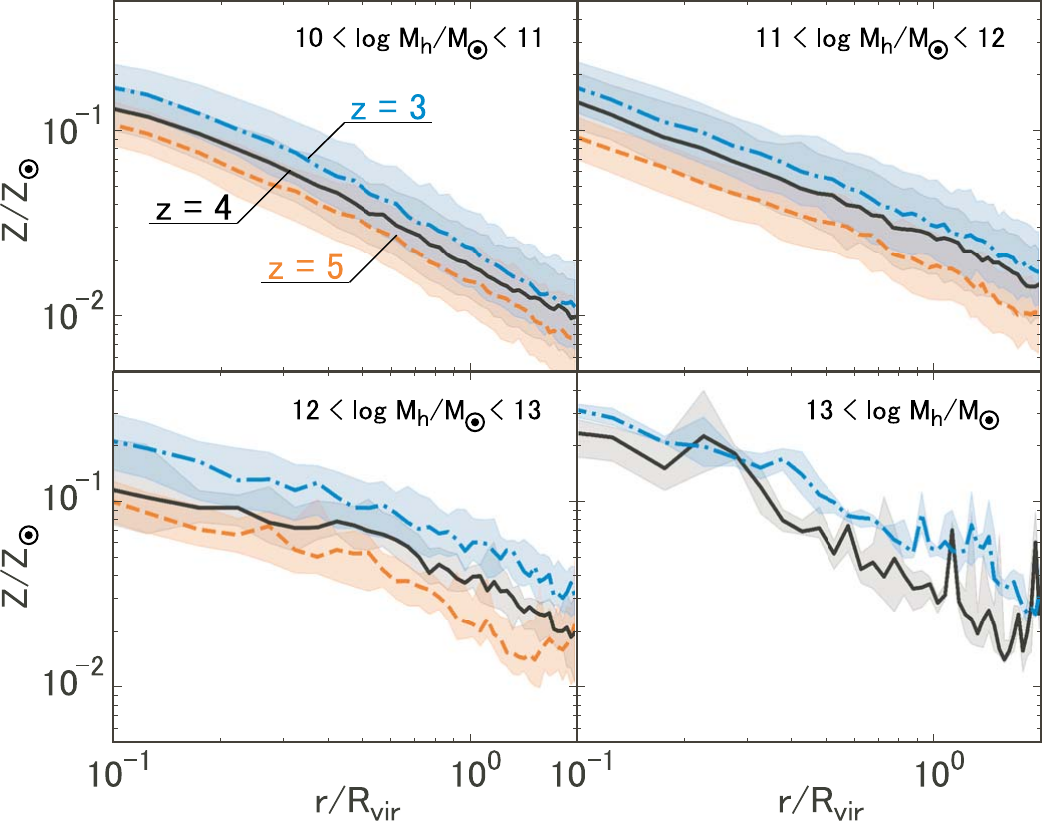}
	\end{center}
	\caption{
	            Radial profiles of gas metallicity. Blue dot-dashed, Black solid and red dashed lines show median values of all haloes in each mass range at $z=3, 4$ and $5$. The shades represent the ranges of 25 - 75 percent.
                }
	\label{fig:Z-r}
\end{figure}

Fig.~\ref{fig:Z-r} presents the radial profiles of metallicities 
 for haloes in PCR0 region from $z = 5.0$ to $3.0$. 
The metallicity monotonically increases with the cosmic time, indicating the accumulation of metals in the CGM region, and also decreases as the radial distance increases. 
 First, metals are produced near the centre of the galaxy and then carried to CGM and IGM via the stellar/AGN feedback \citep{Boissier99}. 
We make fitting as  $ \frac{Z}{\Zsun} \propto \left( \frac{r}{\Rvir} \right)^{\alpha}$ for $0.5 < r/\Rvir < 1$. The slope becomes shallower with increasing halo mass, and it is $-1.06$ for $10^{10} < \Mh/\Msun < 10^{11}$, $-0.85$ for $10^{11} < \Mh/\Msun < 10^{12}$, $-0.77$ for $10^{12} < \Mh/\Msun < 10^{13}$ and $-0.76$ for $10^{13} < \Mh/\Msun$. 
The shallower slopes of massive haloes may reflect that metal outflows stall near virial radii and accumulate into halo gases as suggested in the previous section. 
The distributions of massive haloes with $\Mh > 10^{12}~\Msun$ get shallower as the redshift decreases, while that for lower halo masses does not change significantly with time. This reflects the different efficiencies of the metal accumulation in the CGM depending on the nature of the outflow gas, i.e., the fountain flow or escaping from host haloes.
We note that the AGN-driven outflow would have the potential to affect the metal distribution. For instance, \citet{Taylor+15} argued that the powerful AGN-driven outflow can remove the gas and metals from galaxies and enhance the metallicity of CGM compared to the result without AGN feedback. 
As we mentioned in \S\ref{sec:3.1}, 
BHs grow rapidly and give feedback to surrounding gas once the halo mass exceeds $\sim 10^{12}~\Msun$ in our simulations.
Therefore, the radial profile would also be shallower if the AGN-driven outflow activates in the low-redshift high-mass galaxies and reduces the metals at the central region, though the impact might strongly depend on the modeling of AGN feedback. 
The metallicity profiles are also related to the mass-metallicity relation (MZR) suggested in observations \citep[e.g.,][]{Maiolino08, Mannucci09} and simulations \citep[e.g.,][]{Brooks07, Ma16, Collacchioni20}.
At $r = \Rvir$ and 
$z = 3.0$, for example in our results, $Z/\Zsun$ increases from $0.023$ at $10^{10} < M/\Msun < 10^{11}$ 
(upper-left panel) to $0.061$ at $10^{12} < \Mh/\Msun < 10^{13}$ (lower-left panel), 
but decreases to $0.055$ in massive haloes with $10^{13} < \Mh/\Msun$ (lower-right panel). 
Such a peaking behavior in the MZR is also 
suggested for galaxies at $z \lesssim 1$\citep[]{Collacchioni20}.

The radial metal distribution can be investigated via observations of absorption lines in SEDs of background objects. However, observational sight-lines are too sparse to probe the radial distributions because the background sources are frequently limited to bright 
QSOs due to sensitivities of current observational facilities \citep[e.g.,][]{Steidel10, Lehner14}.
Recently, \citet{Mendez-Hern22} showed the metal distributions from stacking spectra using background galaxies. Future observational missions like the Subaru PFS will be able to detect metal absorption lines in SEDs of many background galaxies and allow us to study the radial metal distributions. With the upcoming data, we will do comparison studies and investigate reasonable star formation and feedback models in future work. 
Note that, there are dispersions of the metal distributions in the same mass range as in the figure. Therefore, statistical studies would be required.

%
%



%
%

\section{Summary} 
In this paper, we investigate gas and metal outflows from massive galaxies in protocluster regions by using the results of the FOREVER22 simulation project that follows the formation of protocluster regions at $z \ge 2$.
Our simulations contain very massive haloes of $\gtrsim 10^{13}~\Msun$ even at high redshifts $z \gtrsim 2$.
Also, there are starburst galaxies with ${\rm SFR} \gtrsim 1000~\Msunyr$ and supermassive black holes with $\Mbh \gtrsim 10^{8}~\Msun$ in the protocluster regions at $z \sim 3$. Here we study the relation between the outflows and the halo mass or SFR, or the gas accretion rate onto a BH. 
The main conclusions are summarized as follows.

\begin{itemize}
    \item The gas outflow rate increases with the halo mass and it reaches $\sim 1000~\Msunyr$ for $\Mh \gtrsim 10^{13}~\Msun$. The mass loading factor decreases with the halo mass, it is $\eta \sim 1 (10)$ for $\Mh \sim 10^{13} (10^{11})~\Msun$.
    \item Massive haloes with $\Mh > 10^{12.5}~\Msun$ show the radial profiles of the gas outflow rates that decrease significantly at virial radii. Therefore the outflow from star-forming regions contributes to halo gas or large-scale fountain flow in the case of massive haloes.
    \item The metal outflow depends on the halo mass and redshift. Considering the metal inflow and outflow, total metal masses in massive galaxies of $\Mh \gtrsim 1 \times 10^{13}~\Msun$ at $z=3$ increase with time, while lower-mass haloes can lose metal via galactic winds.
\end{itemize}

Thus, we suggest the origin of CGM with metals sensitively depends on the halo mass. Future observational missions with James Webb Telescope or the Prime Focus Spectrograph on the Subaru telescope will be able to investigate spatial distributions of gas and metal around high-redshift galaxies. 

%
%
\section*{Acknowledgments}

The numerical simulations were performed on the computer cluster, XC50 in NAOJ, and 
Trinity at Center for Computational Science in University of Tsukuba.
This work is supported in part by MEXT/JSPS KAKENHI Grant Number 17H04827, 20H04724, 21H04489, 22H00149, JST FOREST Program, Grant Number JPMJFR202Z, ABC project research, Grant Number AB041008 (HY).

\section*{Data availability}

The data underlying this article will be shared on reasonable request to the corresponding author.

%
%
\bibliographystyle{mnras}
\bibliography{NH}


\label{lastpage}

\end{document}